\documentclass[runningheads]{llncs}
\usepackage[T1]{fontenc}
\usepackage{graphicx}
\usepackage{listings}
\usepackage{amsmath}
\usepackage{amsfonts}
\usepackage[ruled,linesnumbered]{algorithm2e}
\usepackage{subcaption}
\usepackage[most]{tcolorbox}
\usepackage{booktabs}
\usepackage{hyperref}

\newcommand{\K}{\mathcal{K}}
\newcommand{\Lg}{\mathcal{L}}
\newcommand{\Pl}{\Pi}
\newcommand{\entails}{\vdash}

\begin{document}
\title{Compliance as a Trust Metric}

\author{Wenbo Wu\orcidID{0009-0002-3937-0124} \and
George Konstantinidis\orcidID{0000-0002-3962-9303}}
\institute{University of Southampton, Southampton, UK \\ 
Contact: \href{wenbo.wu@soton.ac.uk}{wenbo.wu@soton.ac.uk}}

\maketitle              
\begin{abstract}
Trust and Reputation Management Systems (TRMSs) are critical for the modern web, yet their reliance on subjective user ratings or narrow Quality of Service (QoS) metrics lacks objective grounding. Concurrently, while regulatory frameworks like GDPR and HIPAA provide objective behavioral standards, automated compliance auditing has been limited to coarse, binary (pass/fail) outcomes. This paper bridges this research gap by operationalizing regulatory compliance as a quantitative and dynamic trust metric through our novel automated compliance engine (ACE). ACE first formalizes legal and organizational policies into a verifiable, obligation-centric logic. It then continuously audits system event logs against this logic to detect violations. The core of our contribution is a quantitative model that assesses the severity of each violation along multiple dimensions, including its \textit{Volume}, \textit{Duration}, \textit{Breadth}, and \textit{Criticality}, to compute a fine-grained, evolving compliance score. We evaluate ACE on a synthetic hospital dataset, demonstrating its ability to accurately detect a range of complex HIPAA and GDPR violations and produce a nuanced score that is significantly more expressive than traditional binary approaches. This work enables the development of more transparent, accountable, and resilient TRMSs on the Web.

\end{abstract}

\section{Introduction}
\label{sec:introduction}
Trust is the foundation that enables interactions among unfamiliar users in distributed networks~\cite{su2015reliable,wu2025trust} like data markets~\cite{fernandez2020data,nguyen2025blockchain,ma2024model}. To facilitate this, TRMSs act as intermediaries that collect and process trust signals to calculate a reputation for each network node, enabling more reliable interaction experiences~\cite{josang2007survey,wu2025trust}. However, traditional TRMSs typically rely on subjective user ratings or context-specific QoS metrics~\cite{wahab2015survey,yu2020crowdr,ghose2007auditing}. 
In parallel, the digital landscape is increasingly governed by formal regulations like GDPR~\cite{gdpr_2018} and HIPAA~\cite{hipaa}, which provide an objective and authoritative foundation for evaluating an entity's compliance level~\cite{truong2019gdpr}. In regulated industries, demonstrable compliance can be the foundational proxy for trustworthiness.
Despite this, the systems we use to measure digital trust have been slow to adapt, leaving compliance auditing as a research gap in modern trust evaluation~\cite{wu2025trust}.

\begin{figure}[t]
    \centering
    \includegraphics[width=\linewidth]{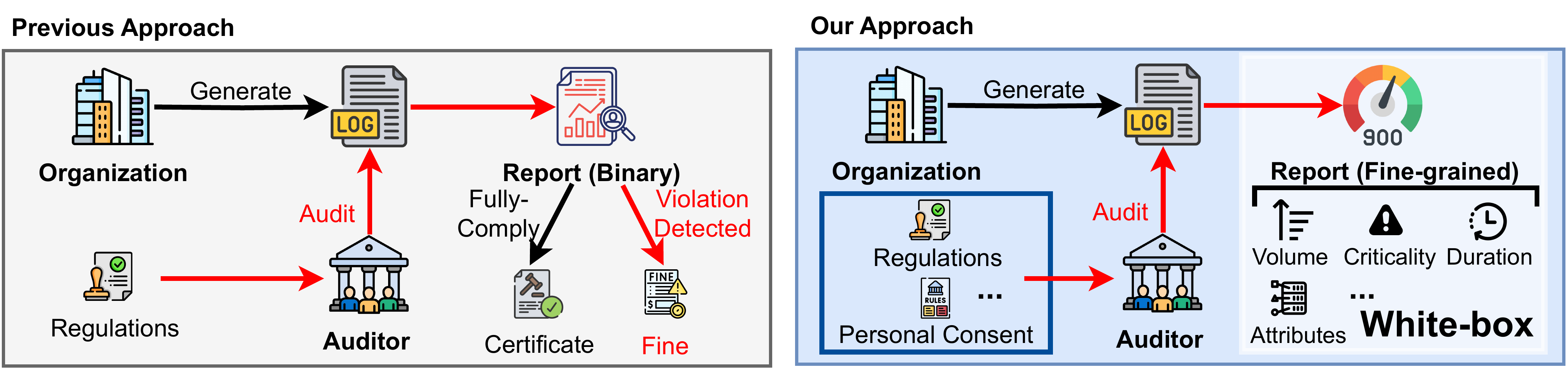}
    \caption{Binary Auditing vs. Our Fine-grained Auditing}
    \label{fig:methods_comparison}
    \vspace{-0.3cm}
\end{figure}

Organizations generate data usage logs to record system operations, and authorities typically audit these logs manually to detect violations. This reliance on manual auditing is problematic, as the process is time-consuming, costly, and error-prone~\cite{9505224}. While automated compliance checking has emerged as a field of study, its methods are incapable of dynamic trust assessment. Existing work (\textit{e.g.}, \cite{9505224,5467075,277104}) predominantly produces binary outcomes—an entity is either compliant or not—which fails to capture the nuanced spectrum of adherence (see the left half of Fig.~\ref{fig:methods_comparison}).
Furthermore, these audit results are typically used for static, single-point-in-time reporting rather than as a live, evolving metric for long-term reputation tracking. This leaves their potential to enrich trust models largely underexplored.

To bridge this gap, we argue that regulatory compliance should be treated not as a binary check, but as a quantitative and dynamic trust metric. 
Accordingly, this paper introduces a novel "white-box" (\textit{i.e.}, the use of explicit, human-readable logic rules rather than opaque neural network embeddings) Automated Compliance Engine (ACE) that provides fine-grained, automated compliance assessment (see the right half of Fig.~\ref{fig:methods_comparison}). This ensures that every generated compliance score is backed by a verifiable, deterministic audit trail, allowing auditors to trace exactly why a specific action was flagged as a violation. ACE first translates unstructured legal regulations into verifiable logic, which enables a continuous audit of system event logs to detect violations. It then quantifies the degree of adherence along multiple dimensions, producing a nuanced compliance score. Finally, this score is designed for seamless integration as an authoritative new dimension within a TRMS, enriching the overall trustworthiness evaluation. ACE has broad applicability in domains, including but not limited to data markets~\cite{fernandez2020data,nguyen2025blockchain,ma2024model} and decentralized finance~\cite{zetzsche2020decentralized,bodo2024trust}, offering a more robust and transferable standard for establishing trust in digital environments.
Our primary contributions are:
\begin{itemize}
    \item \textbf{A Novel Framework for Compliance-Based Trust:} We introduce ACE, a white-box framework that operationalizes regulatory compliance as a verifiable and dynamic trust metric for TRMSs (Sec.~\ref{sec:framework}).
    \item \textbf{Formalization of Policies into Verifiable Logic:} We present a methodology to translate complex legal and organizational policies into a machine-verifiable, obligation-centric logic, enabling automated and continuous auditing (Sec.~\ref{subsec:logic},~\ref{subsec:semantics},~\ref{subsec:violation}, and Appendix~\ref{sec:appendix_methodology_example} and~\ref{sec:appendix_rules}).
    \item \textbf{A Multi-Dimensional Quantitative Scoring Model:} We design a quantitative model that transforms discrete violation data into a fine-grained, continuous compliance score by assessing multiple dimensions of violations, including its volume, duration, breadth, and criticality (Sec.~\ref{subsec:quantifying},~\ref{subsec:scoring}).
    \item \textbf{An Empirical Evaluation of Feasibility and Effectiveness:} We implement and evaluate an open-source prototype of our framework on a synthetic hospital dataset, demonstrating its accuracy in detecting violations and the superior expressiveness of its scoring model compared to traditional binary approaches (Sec.~\ref{sec:experiments}).
\end{itemize}

\section{System Overview}
\label{sec:framework}

\begin{figure}[t]
    \centering
    \includegraphics[width=0.7\linewidth]{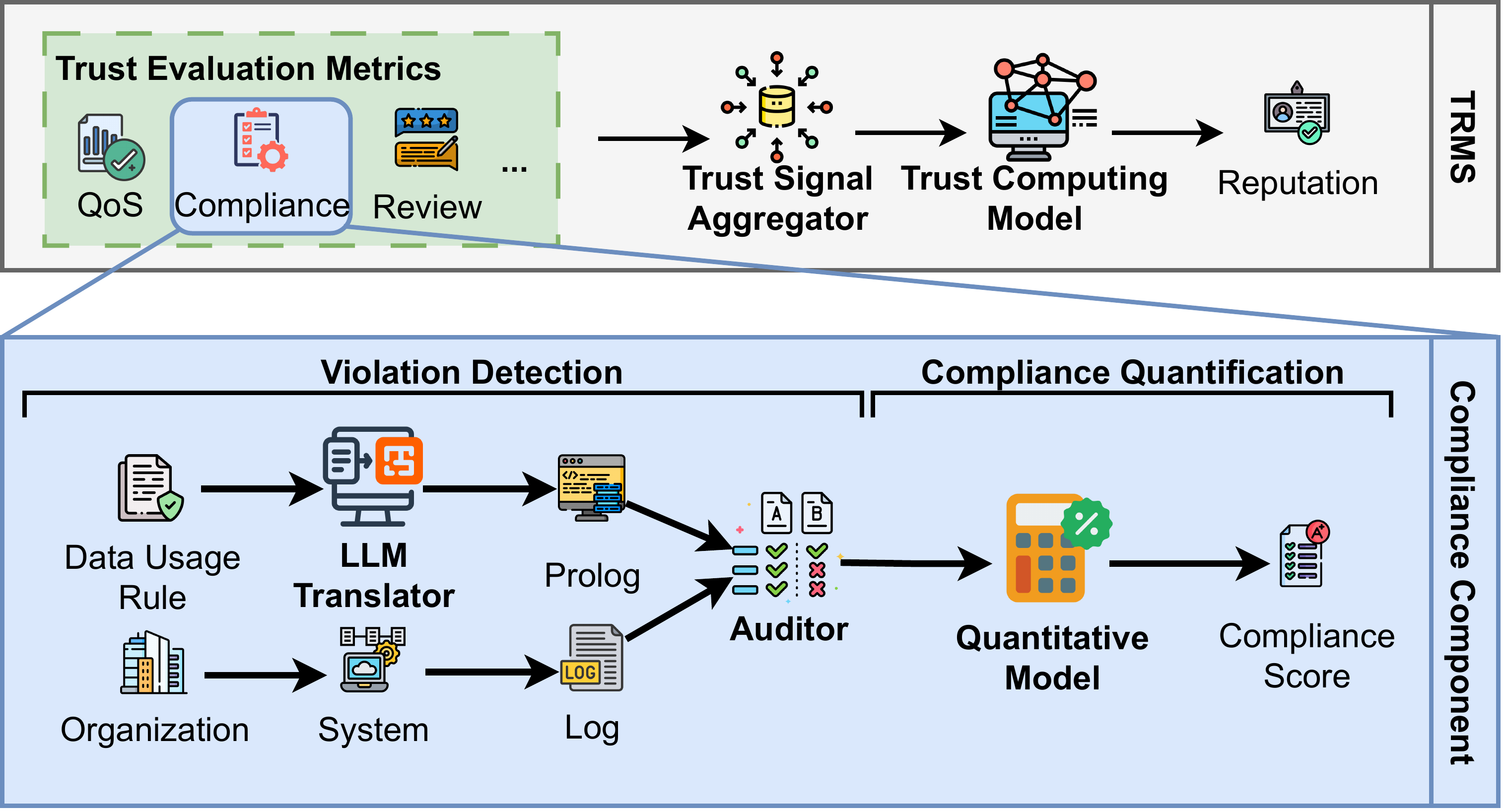}
    \caption{Fine-Grained Compliance Component for TRMS}
    \label{fig:TRMS design}
    \vspace{-0.3cm}
\end{figure}

The overall TRMS design incorporates compliance as a key, quantifiable trust metric, as illustrated in Fig.~\ref{fig:TRMS design}. The framework is designed to generate a holistic reputation score from diverse trust signals, which are classified as explicit (\textit{e.g.}, textual Reviews and ratings) and implicit (\textit{e.g.}, QoS and, crucially, Compliance)~\cite{wu2025trust}. These varied signals are continuously collected and processed by a \texttt{Trust Signal Aggregator}, which applies strategies like time-decay weighting to prepare the data. The prepared data is then fed into a \texttt{Trust Computing Model}, which uses statistical or machine learning algorithms to calculate a final, unified reputation score~\cite{wang2020survey}. This score serves as a key reference, enabling users to make informed, evidence-based decisions when engaging with other entities.

The primary innovation of this framework lies in its fine-grained compliance component, which transforms compliance from a simple ``checkbox'' into a verifiable, data-driven trust metric. Its operation is divided into two main phases: Violation Detection and Compliance Quantification. The process begins by preparing two key inputs. First, all relevant data handling actions performed by an \texttt{Organization} are captured in an immutable system \texttt{Log}, which serves as objective evidence of its behavior. Second, the governing \texttt{Data Usage Rules}, derived from sources like GDPR~\cite{garg2011policy}, patient consent agreements~\cite{konstantinidis2021enabling}, and HIPAA-based policies~\cite{pereira2006role,byun2008purpose,servos2017current}, are formalized. Our framework leverages an \texttt{LLM Translator} to convert these high-level policies into machine-executable \texttt{Prolog} rules.

In the Violation Detection phase, an automated \texttt{Auditor} systematically compares the event \texttt{Log} against the formalized \texttt{Prolog} rules to identify any discrepancies or violations. The audit's output---a stream of identified violations---is then passed to the Compliance Quantification phase. Here, a \texttt{Quantitative Model} assesses these violations based on predefined criteria, such as their frequency and the volume of data affected, to calculate a final numerical \texttt{Compliance Score}.

By processing real-world operational data against formal rules, this component provides a robust and objective measure of an entity's adherence to its commitments. The fine-grained compliance score is then integrated into the main TRMS as a key trust metric, directly linking an entity's reputation to its actual compliance posture. Ultimately, this makes the overall TRMS more objective, transparent, and resilient to manipulation compared to systems that rely solely on subjective feedback.

\section{The Compliance Component}
\label{sec:compliance_component}
At the core of our system is the compliance component (\textit{i.e.}, ACE) formally detecting and quantifying policy violations. This section details the theoretical ground of ACE: (1) the \textit{Compliance Policy Logic} that defines the language of our rules; (2) the \textit{Compliance Verification Semantics} that define the meaning of compliance; (3) the \textit{Compliance Model and Violation Semantics}; (4) the \textit{Quantifying Compliance Violations}; and (5) the \textit{Computing Compliance Score} model that produces a fine-grained compliance score.

\subsection{Compliance Policy Logic}
\label{subsec:logic}
The foundation of our compliance framework is a policy language grounded in a decidable fragment of first-order logic~\cite{9505224,garg2011policy}. This choice provides an expressive, unambiguous syntax for codifying complex, real-world rules. Unlike traditional authorization logics that focus on specifying permissions (i.e., what a user \textit{may} do)~\cite{9505224}, our logic is tailored to express \textbf{obligations} and \textbf{constraints}~\cite{breaux2006towards} (i.e., what \textit{must} be true in a given situation). 
The goal is to formalize policies into verifiable logic to automate compliance auditing and build quantifiable trust.
Following~\cite{9505224,garg2011policy}, the formal syntax of the policy language is defined as:

\begin{definition}[Policy Language Syntax]
\label{def:policy_syntax}
The grammar of our compliance policy language is formally defined by the components:
\begin{align*}
    \text{(Types)} \quad \sigma &::= \texttt{principal} \mid \texttt{resource} \mid \texttt{role} \mid \dots \\
    \text{(Constants)} \quad c &::= \textit{Alice} \mid \textit{Bob\_PHI} \mid \textit{Doctor} \mid \dots \\
    \text{(Variables)} \quad x, y, z &::= \text{A set of typed variables } \mathcal{V} \\
    \text{(Terms)} \quad t &::= c \mid x \quad \text{where } c \text{ and } x \text{ are of the same type } \sigma \\
    \text{(Predicates)} \quad P &::= \texttt{has\_role} \mid \texttt{is\_doctor\_of} \mid \texttt{read} \mid \dots \\
    \text{(Atoms)} \quad \alpha &::= P(t_1, \dots, t_n) \\
    \text{(Formulas)} \quad \phi, \psi &::= \alpha \mid \phi_1 \land \phi_2 \quad \text{(Conjunctive Formulas)} \\
    \text{(Rule)} \quad \rho &::= (\forall \mathbf{x}. (\phi \supset \psi), C) \\
    \text{(Policy)} \quad \Pi &::= \{\rho_1, \dots, \rho_m\}
\end{align*}
\end{definition}

\textit{Types, Terms, and Predicates.}
\textbf{Types} ($\sigma$) partition entities into distinct domains like \texttt{principals} (users, services) and \texttt{resources} (data, objects). \textbf{Terms} ($t$) are the fundamental arguments in logical statements, representing either specific \textbf{constants} ($c$) like the user \textit{Alice}, or \textbf{variables} ($x, y$) that stand for any entity of a given type. \textbf{Predicates} ($P$) declare named relations over these terms, such as $\texttt{is\_doctor\_of}(p, q)$.

\textit{Atoms and Formulas.}
An \textbf{atom} ($\alpha$) is the application of a predicate to a sequence of terms, forming the most basic provable statement (\textit{e.g.}, $\texttt{has\_role}(\textit{Alice},\\ \textit{Doctor})$). In our framework, \textbf{formulas} ($\phi, \psi$) are constructed as conjunctions ($\land$) of these atoms. This structure is expressive enough for a wide range of practical compliance rules while maintaining desirable computational properties similar to Datalog~\cite{huang2011datalog}.

\textit{Compliance Rules: Triggers and Constraints.}
The core of our language is the \textbf{compliance rule} ($\rho$). A policy $\Pi$ is a finite set of such rules. Each rule is a tuple containing a logical formula and a numeric \textbf{Rule Criticality} ($C \in [0, 1]$). The formula itself, $\forall \mathbf{x}. (\phi \supset \psi)$, establishes an obligation:
i) The \textbf{universal quantifier} $\forall$ (read ``for all'') binds the variables in the vector $\mathbf{x}$, making the rule a general statement that applies universally. For brevity, we often assume variables are universally quantified over the scope of the entire rule.
ii) The \textbf{Trigger Condition} ($\phi$) is a formula that acts as the rule's premise. It describes a specific event or state, such as an action being performed. When facts corresponding to the trigger are observed, the rule's obligation is invoked.
iii) The \textbf{implication symbol} `$\supset$' separates the trigger from the constraint.
iv) The \textbf{Required Constraint} ($\psi$) is a formula describing a condition that \textit{must} hold true whenever the trigger is satisfied.

For instance, a hospital policy rule can be: 
$(\forall p_1, p_2, r. (\texttt{read}(p_1, r) \land \texttt{is\_phi}(r) \supset \texttt{has\_role}(p_1, \textit{doctor}) \land \texttt{is\_doctor\_of}(p_1,p_2) \land \texttt{owns\_phi\_record}(p_2,r)), 0.9)$. This rule does not grant permission to read; instead, it asserts that \textit{if} a read on PHI occurs (trigger), then the entity performing the read must have the `doctor' role and is the doctor of the resource owner (constraint). A failure to meet this constraint is a compliance violation.

\subsection{Compliance Verification Semantics}
\label{subsec:semantics}
The verification semantics give formal meaning to our policy language by defining how the truth of a formula is derived from a set of known facts. This process, known as entailment, provides the mechanism for checking the trigger and constraint conditions of a compliance rule. We define this through an \textbf{entailment judgment}:$\K, l \entails \phi$.
This judgment is read as: "The formula $\phi$ is entailed (or is provably true) from the knowledge base $\K$ in the context of a log entry $l$." The log entry $l = (\alpha_{req}, \tau)$ is crucial as it provides the timestamp $\tau$ of the event being verified, allowing for temporally-aware fact checking.

\begin{definition}[Semantic Inference Rules]
\label{def:semantic_rules_revised}
The entailment relation $\entails$ is defined as the smallest relation closed under the following inference rules. These rules specify how to prove positive, conjunctive formulas, which is the foundation for verifying our compliance rules.

\noindent
\textbf{Axiom for Facts.} This rule is the bridge between the stored data and the logic. It states that a ground atom is provably true if a corresponding fact exists in the knowledge base $\K$ and its timestamped validity interval $[T_{start}, T_{end}]$ contains the timestamp $\tau$ of the event being checked.
    \begin{align*}
    \frac{
    (P(c_1, \dots, c_n), T_{start}, T_{end}) \in \K \quad l = (\dots, \tau)  \quad T_{start} \le \tau \le T_{end}
    }{
     \K, l \entails P(c_1, \dots, c_n)
    }   \quad \textnormal{(AXIOM)}
    \end{align*}

\noindent
\textbf{Conjunction Introduction.} This rule defines the semantics of the logical AND ($\land$). It states that a conjunction of formulas is true if, and only if, a proof can be derived for each individual formula. This rule is used to verify the multi-part conditions often found in rule triggers and constraints.
    \begin{align*}
    \frac{
        \K, l \entails \phi_1 \quad \quad \dots \quad \quad \K, l \entails \phi_n
    }{
        \K, l \entails \phi_1 \land \dots \land \phi_n
    }
    \quad \textnormal{(AND-INTRO)}        
    \end{align*}
\end{definition}

\noindent
\textbf{Universal Instantiation ($\forall$-Elimination).}
This rule is the bridge between a general policy and a specific, observable event. It states that if a universally quantified formula is provably true, then any specific instance of that formula, created by substituting the variable with a constant \textit{c} from the system's domain, is also provably true.
\[
    \frac{\mathcal{K},l \vdash \forall x.\phi(x)}{\mathcal{K},l \vdash \phi(c)} \quad (\text{$\forall$-ELIM})
\]

\noindent
\textbf{Implication Elimination ($\supset$-Elimination).}
Commonly known as Modus Ponens, this rule allows the system to derive new facts. If a rule's premise \(\phi\) is proven true, and the rule itself (\(\phi \supset \psi\)) is established, then the conclusion \(\psi\) can be inferred as a new fact. While not strictly required for our violation semantics (Def. 3.4), it enables rule chaining and a more powerful reasoning engine.
\[
    \frac{\mathcal{K},l \vdash \phi \supset \psi \quad \mathcal{K},l \vdash \phi}{\mathcal{K},l \vdash \psi} \quad (\text{$\supset$-ELIM})
\]

These semantic rules provide the formal, operational engine for automated compliance verification. To check a specific action recorded in a log entry $l$ against a rule, the system uses this proof system. As detailed in the next subsection, a violation is detected if, for some substitution $\theta$, the system can successfully build a proof for the trigger condition ($\K, l \entails \phi\theta$) but fails to build a proof for the required constraint ($\K, l \not\entails \psi\theta$).

\subsection{Compliance Model and Violation Semantics}
\label{subsec:violation}
To quantify compliance, we first need a formal model to define what constitutes a violation. We extend our logical framework from defining permissions to specifying \textbf{obligations} and \textbf{constraints}. A violation, therefore, is not merely an un-permitted action, but an action that breaks a required constraint.

\begin{definition}[Compliance Model]
A system is a tuple $(\K, \Lg, \Pl)$, where:
\begin{itemize}
    \item $\K$ is the \textbf{Knowledge Base}, is the system's central repository of ground-truth information, acting as the authoritative source of facts against which compliance is measured. It is not a static database but is formally defined as a set of timestamped facts.
    \item $\Lg$ is the \textbf{Access Log} is the system's complete, time-ordered, and immutable record of all actions, serving as the objective evidence consumed by the automated Auditor. The conceptual log can comprise a \texttt{staff activity log} for internal operations and a \texttt{patient request log} for external compliance-related actions like data access or erasure requests.
    \item $\Pl$ is the \textbf{Compliance Policy}, a set of compliance rules. Each rule $\rho \in \Pl$ is a pair $(\forall \mathbf{x}. (\phi \supset \psi), C)$, where $C$ is rule criticality.
\end{itemize}
\end{definition}

Unlike the authorization logic which defines what a principal \textit{may} do, this logic defines what \textit{must} be true when a certain action occurs. A violation occurs when the trigger condition is met but the required constraint is not.




        
        

\begin{definition}[Violation Semantics]
\label{def:violation_main}
An access log entry $l = (\alpha_{req}, \tau)$ constitutes a \textbf{violation} of a rule $\rho = (\forall \mathbf{x}. (\phi \supset \psi), C)$ if there exists a ground substitution $\theta$ that aligns the log entry with the rule's trigger, but the rule's constraint is not met. Formally, a violation occurs if:
\begin{equation}
\underbrace{\K, l \entails \phi\theta}_{\text{Trigger is met}} \quad \land \quad \underbrace{\K, l \not\entails \psi\theta}_{\text{Constraint is not met}}
\end{equation}
where the judgment $\entails$ is derived using the enforcement semantics (Sec.~\ref{subsec:semantics}). Typically, the trigger formula $\phi$ includes an action atom that unifies with $\alpha_{req}$ from the log entry.
\end{definition}

\begin{theorem}[Violation as Logical Inconsistency]
The detection of a violation for a log entry $l$ under a policy $\Pl$ is equivalent to finding a logical inconsistency between the compliance rule $\rho \in \Pl$ and the observed state of the world (i.e., the log entry $l$ combined with the facts in $\K$).
\end{theorem}
\textit{Proof Sketch:}
The proof follows from the definition of a violation (Def. \ref{def:violation_main}). The detection algorithm finds a substitution $\theta$ such that the premise of an implication ($\phi\theta$) is true, while the conclusion ($\psi\theta$) is false. This configuration, $\phi\theta \land \neg\psi\theta$, is a direct contradiction of the rule's assertion that $\phi\theta \supset \psi\theta$ must hold for all substitutions. The algorithm is a direct implementation of this semantic check, ensuring no false positives. \qed

\subsection{Quantifying Compliance Violations}
\label{subsec:quantifying}
A binary violation flag lacks the granularity needed for a dynamic trust metric. To capture the true magnitude of non-compliance, we introduce a quantitative model that operates over discrete, configurable time windows, $W$ (\textit{e.g.}, weekly or monthly). This periodic approach assesses a principal's behavior within a given time window, providing a basis for a continuously evolving compliance score.

\noindent
\textbf{Periodic Violation Metrics}
For each principal $p$ and each time window $W$, all detected violations are first grouped by the specific rule $\rho_i \in \Pi$ that was broken. From these groupings, we derive a set of raw metrics that characterize the scope and persistence of the non-compliant behavior for that rule.

\begin{definition}[Per-Rule Violation Metrics]
\label{def:violation_metrics}
For each rule $\rho_i$ violated by a principal $p$ during a window $W$, we compute the following metrics:
\begin{itemize}
    \item \textbf{Volume ($M_{V}$):} The total number of unique resources (the number of specific rows/records, \textit{e.g.}, 50 patient files) affected by violations of rule $\rho_i$. This captures the scale of the non-compliance. 
    \item \textbf{Duration ($M_{T}$):} The total time elapsed from the first to the last logged violation of rule $\rho_i$ within the window. This measures the persistence of the failure to comply.
    \item \textbf{Breadth ($M_{B}$):} The number of distinct resource attributes (the number of columns/fields affected, \textit{e.g.}, accessing 'DoB' and 'Address' = breadth of 2) involved in violations of rule $\rho_i$. This assesses the scope of the incident across different data categories. 
\end{itemize}
\end{definition}

\noindent
\textbf{Normalizing Violation Metrics}
The raw metrics exist on different, unbounded scales. To combine them meaningfully, each metric is transformed into a normalized score on a $[0, 1]$ scale using a negative exponential function. This function models the principle of diminishing impact, where each additional unit of violation contributes less to the overall severity than the one before it.

\begin{definition}[Normalized Severity Components]
\label{def:severity_components}
For each raw metric $M \in \{M_V, M_T, M_B\}$, its normalized score $S$ is given by:
$$ S(M) = 1 - e^{-(M / \alpha)} $$
where $\alpha$ is a configurable scaling parameter unique to each metric ($\alpha_V, \alpha_T, \alpha_B$). This parameter defines the characteristic value at which the severity is considered significant.
A small $\alpha$ denotes high sensitivity to minor violations, while a large $\alpha$ requires a greater violations to yield a high score.
\end{definition}

\noindent
\textbf{Calculating Violation Magnitude}
With normalized components, we can compute a unified magnitude score for each rule violation. We employ a weighted geometric mean to capture the synergistic effect of multiple dimensions of non-compliance.

\begin{definition}[Per-Rule Violation Magnitude]
The Magnitude $M_{i,p,W}$ for a rule $\rho_i$ violated by principal $p$ in window $W$ is the weighted geometric mean of its normalized severity components, this is theoretically grounded in the "synergistic effect" of non-compliance:
$$ M_{i,p,W} = \left( S_V^{w_V} \cdot S_T^{w_T} \cdot S_B^{w_B} \right) $$
where the weights $w_j \ge 0$ and $\sum w_j = 1$. 
We employ a weighted geometric mean rather than a linear sum to mathematically enforce that "widespread, persistent, and broad violations are disproportionately more severe than the linear sum of their parts".
This ensures that a low score in any single dimension significantly dampens the overall magnitude, preventing the system from over-penalizing trivial incidents while accurately capturing the scale of complex violations. 
\end{definition}

\subsection{Computing the Compliance Score}
\label{subsec:scoring}
The final stage of our model aggregates the per-rule violation data into a single, evolving compliance score for each principal. This is achieved by first calculating a total severity penalty for the current period, and then updating a long-term, time-decaying cumulative penalty.

\noindent
\textbf{Per-Period Severity}
First, the magnitude of each rule violation is scaled by the rule's intrinsic importance. These scaled values are then summed to find the total penalty incurred during the period.

\begin{definition}[Total Per-Period Severity]
\label{def:total_severity}
The total severity score $Sev_{p, W}$ for a principal $p$ in a window $W$ is the sum of the severity scores of all rules they violated. The severity of a single rule violation is its magnitude multiplied by its predefined criticality $C_i \in [0, 1]$.
$$ Sev_{p, W} = \sum_{i \in \text{violated rules}} (C_i \cdot M_{i,p,W}) $$
This additive approach is justified as violations of distinct rules represent independent and cumulative failures of compliance.
\end{definition}

\noindent
\textbf{Evolving Principal Compliance Score}
Unlike static, single-window scores, our final compliance score evolves over time, reflecting a principal's entire history while giving more weight to recent actions. This provides a path to redemption for principals who improve their behavior.

\begin{definition}[Compliance Score]
\label{def:compliance_score}
The compliance score for a principal $p$ at the end of period $W_k$ is derived from a time-decaying cumulative penalty. The penalty is updated recursively:
$$ \text{Penalty}_k = (\text{Penalty}_{k-1} \cdot e^{-\lambda}) + Sev_{p, W_k} $$
where $\text{Penalty}_{k-1}$ is the cumulative penalty from the previous period and $\lambda$ is a "decay constant" that determines how quickly past violations are forgiven, which serves as the mathematical realization of a "Path to Redemption".
The final compliance score is then calculated by mapping this unbounded penalty to the range $[0, 1]$ using the hyperbolic tangent function:
$$ Comp(p, W_k) = 1 - \tanh(\text{Penalty}_k) $$
A score of 1 indicates perfect compliance, while a score approaching 0 indicates a history of severe and/or recent non-compliance.
\end{definition}

\begin{theorem}[Boundedness of Score]
The final Compliance Score $Comp(p, W_k)$ is guaranteed to be in the range $[0, 1]$.
\end{theorem}
\textit{Proof Sketch:} The hyperbolic tangent function, $\tanh(x)$, is bounded in the range $[-1, 1]$ for any real input $x$. Since the Penalty score is always non-negative, $\tanh(\text{Penalty}_k)$ is bounded in $[0, 1]$. Consequently, $1 - \tanh(\text{Penalty}_k)$ is also guaranteed to be in the range $[0, 1]$, making it a well-formed input for TRMSs. \qed

\section{Experiments}
\label{sec:experiments}
To validate our proposed framework, we conduct a series of experiments designed to evaluate the accuracy and performance of our auditor, ACE, in detecting a variety of compliance violations within a simulated hospital environment. Also, we demonstrate that our fine-grained quantitative model produces a compliance score that is more expressive and informative in distinguishing the severity of different violations when compared to traditional baseline models.

\subsection{Experimental Setup}
The ACE auditor uses SWI-Prolog 9.0 as its logic engine, integrated via the \texttt{pyswip} library. The auditing and quantitative scoring were performed using  Python 3.9. All experiments were conducted on an Apple machine with an M1 Max chip and 32GB of memory.

\noindent
\textbf{Policy Corpus:}
We formalize five key compliance rules relevant to a hospital scenario, as detailed in Section~\ref{sec:compliance_component} and Appendix~\ref{sec:appendix_rules}. These include three rules based on GDPR data subject rights (Art. 15, 17, 18), one authorization rule and one minimum necessary rule based on the principles of HIPAA.

\noindent
\textbf{Dataset:}
We develop a synthetic data generator designed as an augmented derivation of the MIMIC-III dataset~\cite{johnson2016mimic} structure. This generator produces a hospital dataset comprising a \texttt{Knowledge Base}, a \texttt{Staff Activity Log}, and a \texttt{Patient Request Log} that mirrors the specific attributes, schema, and relational complexity (\textit{e.g.}, patient-doctor assignments, billing codes, and PHI timestamps) found in the MIMIC-III critical care database. While the specific entries are generated to ensure privacy, this alignment ensures that the data's structural complexity is representative of real-world healthcare environments. The dataset includes three types of principals: doctor, patient, and billing clerk. We adopted this augmented approach rather than utilizing raw clinical logs because existing real-world datasets lack the explicit ground-truth violation labels required for precise validation. Our generator is capable of injecting specific, verifiable violation scenarios (\textit{e.g.}, complex GDPR "Right to be Forgotten" failures) with corresponding labels, establishing a rigorous ground truth for our analysis.

\subsection{Evaluation of Violation Detection} \label{subsec:eval-violation-detection}

We evaluate ACE's violation-detection accuracy and performance using the synthetic datasets. Each dataset is constructed with 5\% labeled violating entries under the single-rule-per-violation constraint. The auditor is executed on both staff-activity and patient-request logs across dataset sizes from 5k to 500k.

\noindent
\textbf{Detection accuracy.} Across all 10 runs (five staff logs and five patient logs), the auditor reported exactly the same number of rule instances as the number of labeled violations. This corresponds to perfect recall on the labeled violation set (100\% of labeled violations were detected) and confirms that the ACE auditor can accurately detect all the violations defined in the policy rules.

\begin{table*}[t]
\centering
\caption{Performance of the ACE Auditor.}
\label{tab:violation-detection}
\resizebox{\textwidth}{!}{%
\begin{tabular}{l|lll|lll}
\toprule
{\textbf{Log Entries}} & {\textbf{Runtime}} & {\textbf{Throughput}} & {\textbf{Peak Memory}} & {\textbf{Runtime}} & {\textbf{Throughput}} & {\textbf{Peak Memory}}\\
\midrule
& \multicolumn{3}{c}{\textit{Staff Activity Log}} & \multicolumn{3}{|c}{\textit{Patient Request Log}} \\
\midrule
5,000      & 50.173 s & 99.655 entries/s &  81.66 MB & 57.076 s & 87.603 entries/s & 78.33 MB\\
10,000    & 112.006 s & 89.281 entries/s & 81.5 MB & 112.699 s & 88.732 entries/s & 80.5 MB \\
50,000    & 596.302 s & 83.85 entries/s & 117.11 MB & 659.441 s & 75.822 entries/s & 125.0 MB \\
100,000   & 1299.936 s & 76.927 entries/s & 179.92 MB & 1291.577 s & 77.425 entries/s & 159.62 MB \\
200,000   & 2654.017 s & 75.357 entries/s & 221.16 MB & 2666.071 s & 75.017 entries/s & 231.64 MB \\
500,000   & 6363.378 s & 78.575 entries/s & 413.72 MB & 6688.031 s & 74.76 entries/s & 485.39 MB \\
\bottomrule
\end{tabular}}
\end{table*}

\noindent
\textbf{Scalability and resource use.} Table~\ref{tab:violation-detection} summarizes runtime, throughput (entries per second), and peak resident memory observed for each test. Runtime grows approximately linearly with the number of rows. The patient-request series exhibits similar behavior. Throughput declines modestly with scale indicating a roughly constant per-row processing cost with modest overhead growth. Peak memory usage increases with dataset size but remains comfortably below 500 MB for our largest runs.

\noindent
\textbf{Implications.} The empirical findings show that ACE achieves perfect detection while exhibiting near-linear runtime scaling in the number of log entries. The measured throughput indicate that ACE can handle medium-to-large offline audit workloads on commodity hardware. The primary bottleneck is overall per-row processing time (driven by KB assertions, Prolog query overhead, and I/O). For larger deployments or near-real-time auditing, straightforward optimizations are available: batching assertions, reducing KB re-loading, parallelizing independent record checks, or moving hot predicates to a compiled Prolog module.

\subsection{Evaluation of the Quantitative Scoring}

Beyond the core accuracy of violation detection, we also evaluate the effectiveness and expressiveness of our quantitative scoring model. The following case studies are designed to demonstrate how the fine-grained score provides a more nuanced and dynamic measure of compliance compared to traditional approaches.

\noindent
\textbf{Differentiating Violation Severity}
This experiment evaluates the model's primary claim of expressiveness. The goal is to demonstrate that the ACE produces a nuanced score of non-compliant scenarios that fundamentally differ in traditional models are expected to fail.

We compare the ACE compliance score against two standard baselines: a binary model and a simple count-based model. The evaluation is conducted across four crafted scenarios for a single principal, \texttt{doctor\_A}, within a 30-day window. The baselines are defined as follows:
\begin{itemize}
    \item \textbf{Binary Model:} $S_{\text{binary}}=0$ if any violation is detected, otherwise 1. This model captures only the presence or absence of non-compliance.
    \item \textbf{Count-Based Model:} $S_{\text{count}} = 1 - (N_{\text{violations}}/N_{\text{total}})$, where $N_{\text{total}}$ is the total number of log entries audited. This model's assessment is driven solely by the frequency of violation events.
\end{itemize}

Four scenarios were designed to isolate the impact of different violation types:
\textbf{Scenario A (Low Impact):} A single, isolated violation of a medium-criticality rule (\texttt{hipaa\_auth\_control}, $C=0.8$) affecting one resource for one day.
\textbf{Scenario B (High Volume):} A large-scale violation of the same rule, affecting 50 unique resources over a two-day period.
\textbf{Scenario C (High Criticality):} A single, severe violation of a high-criticality rule (\texttt{gdpr\_art17\_erasure}, $C=0.95$), where a statutory 30-day deadline was missed.
\textbf{Scenario D (High Duration):} A persistent violation involving repeated unauthorized access to a single resource, occurring once per day for 25 consecutive days.

\begin{table}[t]
\centering
\caption{Comparison of model scores across four violation scenarios. A lower compliance score indicates a more severe assessment of the principal's behavior.}
\label{tab:severity_comparison}
\resizebox{0.5\textwidth}{!}{%
\begin{tabular}{l c c c c}
\toprule
\textbf{Scenario} & \textbf{\#violations} & \textbf{ACE} & \textbf{Binary} & \textbf{Count} \\
\midrule
A: Low Impact & 1 & \textbf{0.553} & 0 & 0.99 \\
B: High Volume & 50 & \textbf{0.364} & 0 & 0.50 \\
C: High Criticality & 2 & \textbf{0.478} & 0 & 0.98 \\
D: High Duration & 25 & \textbf{0.420} & 0 & 0.75 \\
\bottomrule
\end{tabular}}
\end{table}

The comparative results, presented in Table~\ref{tab:severity_comparison}, highlight the superior expressiveness of the ACE model.
As anticipated, the Binary model is unable to differentiate between any of the non-compliant scenarios, assigning a score of 0 to all. The Count-Based model, while offering some differentiation, produces a misleading assessment; for instance, it rates the highly critical GDPR violation (Scenario C) as less severe than the persistent access in Scenario D, and nearly identical to the low-impact Scenario A, because it is insensitive to rule criticality.

In contrast, the ACE model provides a more nuanced evaluation. It correctly identifies the single low-impact event (A) as the least severe. It also produces a distinct ranking for the other scenarios based on the current parameterization, which is ordered by severity as:
\(\text{B (High Volume)} \succ \text{D (High Duration)} \succ \text{C (High Criticality)} \succ \text{A (Low Impact)}.\)
This demonstrates that the ACE score is sensitive to multiple dimensions of non-compliance. Furthermore, the model's parameterization allows for ranking to be tuned to reflect specific organizational priorities. For example, by increasing the weight of the criticality component, an organization can configure the model to ensure that Scenario C is rated as the most severe, thereby aligning the quantitative score with its specific risk posture.

\noindent
\textbf{Dynamic Score Evolution}
This case study demonstrates the dynamic, time-aware nature of the ACE score by tracking the evolving compliance of multiple principals over an extended period. The goal is to show how the score provides a nuanced, continuous measure of trustworthiness that is more expressive than simple violation counts. We simulated an eight-month activity log for six principals (one billing clerk and five doctors) and calculated their compliance scores at the end of each month. The goal is to visualize how the compliance score degrades in response to violations and, crucially, recovers when compliant behavior is restored, demonstrating the model's "path to redemption" feature.

\begin{figure*}[t]
    \centering
    \includegraphics[width=\linewidth]{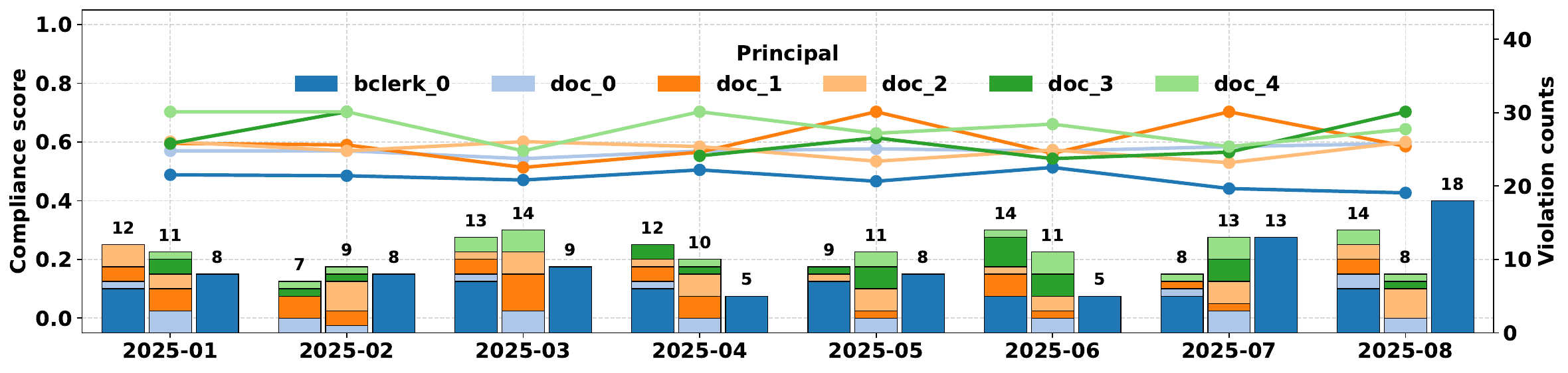}
    \caption{Monthly compliance trends and per-month rule violations}
    \label{fig:score_trends}
    \vspace{-0.3cm}
\end{figure*}

Figure~\ref{fig:score_trends} plots the monthly compliance scores (left axis, line graph) and the raw violation counts (right axis, bar chart) for each principal. The results highlight several key features of our model:
\begin{itemize}
    \item \textbf{Dynamic Responsiveness:} The compliance scores for all principals fluctuate monthly, directly responding to their actions within each period. This confirms the model's ability to serve as a live, dynamic metric for tracking behavior over time.
    \item \textbf{Nuance Beyond Raw Counts:} The score is not a simple inverse of the violation count. For instance, in month 2025-08, \texttt{bclerk\_0} and \texttt{doc\_1} have similar violation counts (14 and 13, respectively), yet their compliance scores differ significantly (approximately 0.4 vs. 0.6). This demonstrates that the ACE score incorporates the severity of violations rather than just frequency.
    \item \textbf{Long-Term Trend Tracking:} The visualization effectively captures long-term behavioral trends. For example, \texttt{doc\_3} shows a general upward trend in their compliance score after an initial dip, while \texttt{bclerk\_0}'s score remains consistently lower, reflecting a persistent pattern of more severe non-compliance. This makes the score suitable for long-term trust and reputation tracking, where historical behavior and possibilities for recovery are essential.
\end{itemize}

\begin{figure}[t]
    \centering
    \begin{subfigure}{0.35\linewidth}
        \includegraphics[width=\linewidth]{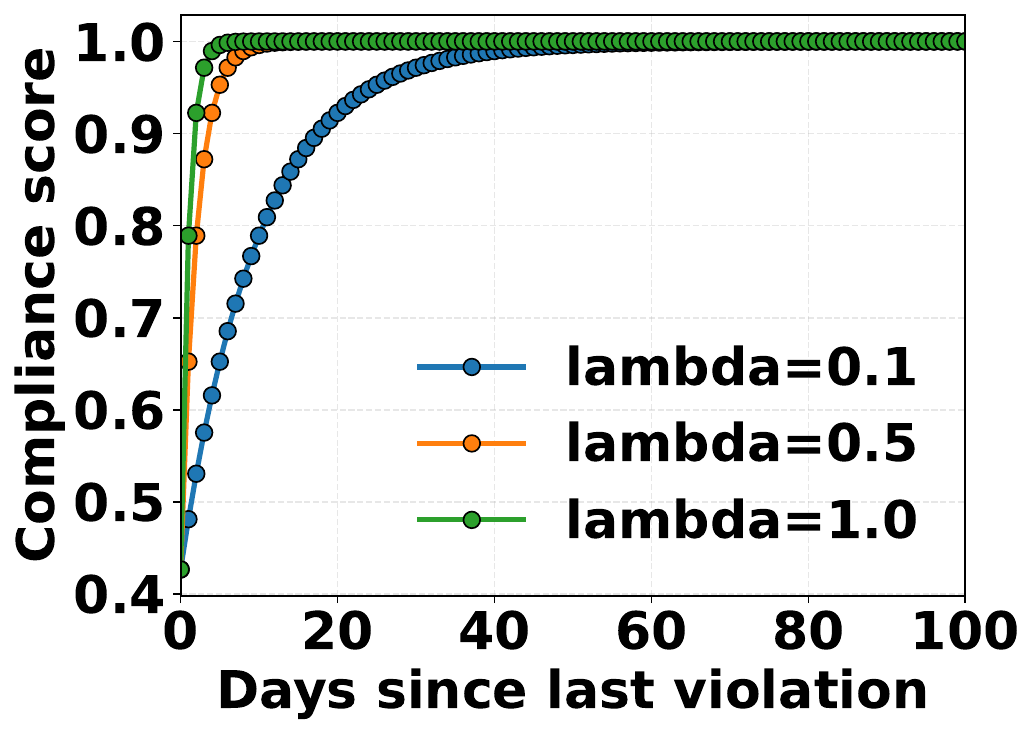}
    \end{subfigure}
    \hfill
    \begin{subfigure}{0.35\linewidth}
        \includegraphics[width=\linewidth]{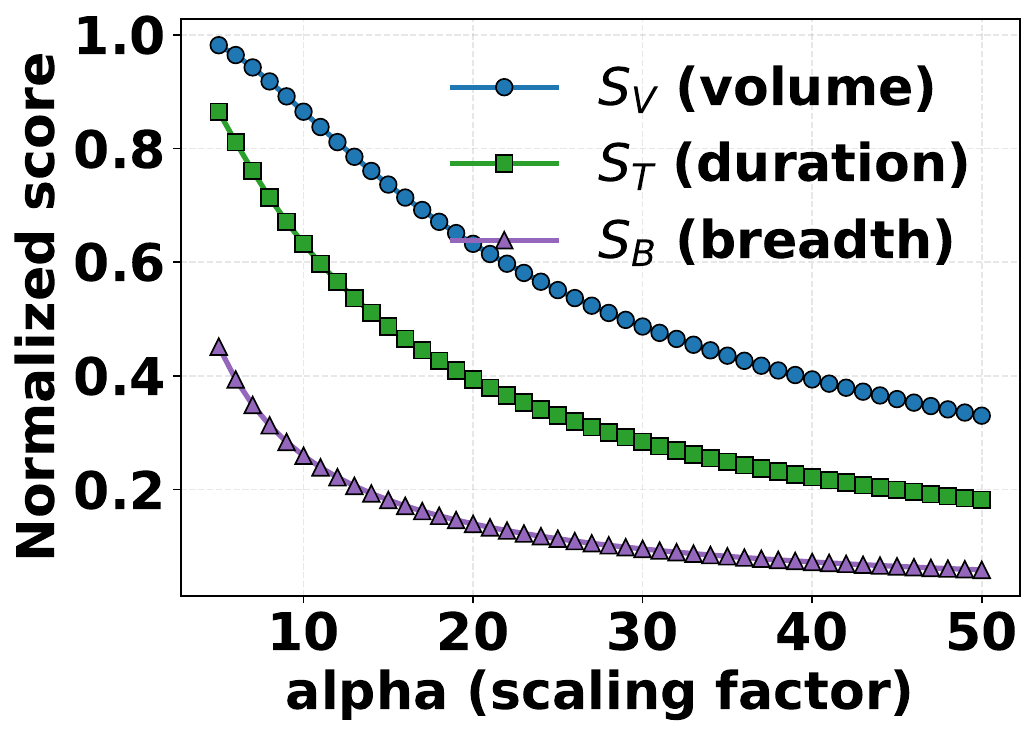}
    \end{subfigure}
    \caption{Parameter sensitivity: $\lambda$ (left) and $\alpha$ (right).}
    \label{fig:sensitivity}
    \vspace{-0.3cm}
\end{figure}

\noindent
\textbf{Sensitivity Analysis of Model Parameters}
This experiment analyzes the impact of the model's key tunable parameters on the final compliance score (the other parameters are analyzed in Appendix~\ref{sec:appendix_system_parameters}), demonstrating its flexibility. The results in Figure~\ref{fig:sensitivity} show how the model can be configured to align with different organizational priorities and risk tolerances by adjusting the decay constant ($\lambda$) and the normalization scaling factors ($\alpha$).

\noindent
\textit{Analysis of the Decay Constant ($\lambda$)}
We analyzed the model's "path to redemption" feature by varying the decay constant $\lambda$, which controls how quickly past violations are forgiven. The left panel of Fig.~\ref{fig:sensitivity} plots the compliance score's recovery over 100 days since the last violation for three different $\lambda$ values. A higher value of $\lambda$ ($\lambda=1.0$) corresponds to high forgiveness, with the score recovering to near-perfection in approximately 10 days. Conversely, a lower value ($\lambda=0.1$) represents low forgiveness, with the score recovering much more slowly. This confirms that the $\lambda$ parameter effectively controls the system's memory and allows it to be tuned to be more or less forgiving of past violations.

\noindent
\textit{Analysis of the Scaling Factor ($\alpha$)}
We then analyzed the sensitivity of the severity calculation to the scaling factor $\alpha$. The right panel of Fig.~\ref{fig:sensitivity} shows the normalized severity score for fixed raw violations ($M_V=20, M_T=10, M_B=3$) as their corresponding scaling factors, $\alpha_V, \alpha_T, \alpha_B$, are adjusted. A small $\alpha$ ($\alpha=5$) results in a high severity score, making the system highly sensitive to even a moderate number of violations. A large $\alpha$ ($\alpha=50$) yields a much lower score, indicating that a greater number of violations would be needed to be considered severe.

\section{Related Work}
\label{sec:related_work}

Our research is positioned at the intersection of three key areas: TRMSs, automated compliance checking, and the quantitative compliance assessment.

\noindent
\textbf{Trust and Reputation Management}
Trust management has a long history in distributed systems~\cite{liu2021behavior,liu2023survey,wu2025trust}. Early work focused on formal, rule-based systems for trust negotiation~\cite{5467075} and managing trust assertions on the Semantic Web~\cite{10.1007/978-3-540-39718-2_23}. Subsequent research integrated these policy-based approaches with reputation mechanisms that aggregate behavioral feedback over time~\cite{bonatti2007integration}. Other related works explore how to measure trust in data-driven systems by discovering conformance constraints~\cite{fariha2021conformance} or how user-facing privacy explanations can foster end-user trust~\cite{BRUNOTTE2023111545}.
While foundational, these systems typically rely on subjective user ratings, pre-negotiated service level agreements, or narrow QoS metrics~\cite{manuel2015trust,goo2008facilitating}. They often lack a continuous, objective, and verifiable signal grounded in adherence to formal regulations. Our work fills this gap by introducing verifiable regulatory compliance as a new, high-integrity input for TRMSs. 

\noindent
\textbf{Automated Compliance Checking}
The field of automated compliance checking focuses on algorithmically verifying system behavior against a set of policies~\cite{truong2019gdpr,wang2023enabling}. Systems like PrivGuard have made significant strides in making privacy regulation compliance easier to check~\cite{277104}. Researchers have also developed formal methods for automating audits, even when system logs are incomplete~\cite{9505224}. Other approaches have leveraged Natural Language Processing (NLP) to create datasets for identifying regulation compliance in unstructured software privacy policies~\cite{zhao-etal-2022-fine-grained,zhang2016semantic,zhou2022integrating}.
However, the primary output of these systems is typically a binary decision: an entity is either compliant or not. These tools are designed for internal audits or enforcement actions rather than for generating a dynamic, public-facing trust signal. Our framework builds upon the outputs of such systems, transforming their discrete violation data into a continuous, fine-grained score suitable for reputation management.   

\noindent
\textbf{Quantitative Compliance Assessment}
The most closely related research shares our motivation for moving beyond binary compliance. Zhang \textit{et al.} proposed a method for quantitative evaluation of compliance levels in the context of safety regulations~\cite{ZHANG201981}. 
Chen \textit{et al.}~\cite{chen2025yesnopredictivecompliance}  proposed the concept of quantifying compliance, their work is strictly limited to predictive monitoring using "black-box" machine learning models to forecast temporal delays (a single dimension) with respect to policy deadlines.
While these works pioneer the concept of quantitative compliance, our contribution is distinct and complementary. As we solve fundamentally different problems using distinct technical architectures.
We introduce a comprehensive, end-to-end framework that begins with a formal, obligation-centric logic for policy definition. We then propose a novel, multi-dimensional quantification model and a time-decaying aggregation mechanism specifically designed to produce an evolving score. Crucially, we are the first to explicitly design this quantitative compliance score as a modular and integrable trust metric for enhancing modern TRMSs.

\section{Discussion}

In this section, we discuss: 1) the practical applicability of the ACE within distributed ecosystems, 2) analyze deployment considerations regarding log integration, and 3) demonstrate the model's robustness against strategic exploitation.

\noindent
\textbf{Use Cases}
While traditional regulatory frameworks often rely on static, infrequent audits (\textit{e.g.}, annual GDPR certifications~\cite{lachaud2020gdpr}), modern distributed environments require real-time operational trust visibility. Continuous scoring adds distinct value in dynamic ecosystems, such as decentralized data markets~\cite{fernandez2020data,ma2025mixture,nguyen2025blockchain}. In these environments, data sellers must establish trust that data buyers will continuously comply with data usage agreements throughout the transaction lifecycle, rather than relying on a compliance snapshot from a previous audit cycle. ACE bridges this gap by providing a live, evolving metric, enabling participants to make decisions based on the current compliance posture of a counterparty. 

\noindent
\textbf{Log Conversion}
A practical concern for deploying ACE is the overhead associated with converting heterogeneous institutional logs into the system's verifiable format. We clarify that this process constitutes a standard Extract-Transform-Load (ETL) task. The integration requires a one-time schema mapping to align internal system logs with ACE's Prolog-based fact structure. Furthermore, the risk of conversion errors or data misalignment is significantly mitigated by the strict typing defined in our \textit{Policy Language Syntax} (Def.~\ref{def:policy_syntax}). This syntax acts as a validation layer, rejecting malformed inputs or type mismatches before the auditing process begins, thereby ensuring the integrity of the evaluation data.

\noindent
\textbf{Robustness Against Strategic Exploitation}
A requirement for any quantitative auditing system is adversarial resilience, where a malicious actor might attempt to tactically violate rules to mask severe issues. The mathematical structure of the ACE model can inherently prevent such exploitation through three key mechanisms:
\textit{i).} Unlike average-based systems where "good" behavior dilutes "bad" grades, ACE utilizes an Additive Penalty Model (Def.~\ref{def:total_severity}). Severity is the sum of weighted violations; thus, committing additional minor violations increases the penalty stack rather than averaging it down.
\textit{ii).} The Rule Criticality parameter ensures high-impact violations dominate the score regardless of lower-impact compliance. This prevents entities from masking critical breaches behind a volume of trivial compliant actions.
\textit{iii).} The system counters "splitting" attacks by grouping related events into single maximal instances based on timestamp (Defs.~\ref{def:violation_metrics},~\ref{def:severity_components}). Furthermore, recovery relies solely on Time Decay (Def.~\ref{def:compliance_score}); entities cannot recover reputation with compliant actions but must cease violations to rebuild trust over time. 

\section{Conclusion}
In this paper, we introduced ACE, a novel white-box framework that transforms regulatory compliance from a static, binary assessment into a quantitative and dynamic trust metric. By formalizing legal and organizational policies into a verifiable, obligation-centric logic, our system continuously audits event logs to compute a fine-grained compliance score. This score, derived from multiple dimensions, assessing violation volume, duration, breadth, and criticality, offers an expressive measure of trustworthiness than traditional approaches.

While our evaluation demonstrated the ACE's accuracy and the nuanced output of our quantitative model, we acknowledge several limitations that pave the way for future research. First, the process of translating complex, often ambiguous legal text into formal logic currently requires significant manual effort from domain experts. Second, our validation was conducted on a synthetic dataset; testing the ACE's robustness against the noise and inconsistencies of real-world operational logs is a vital next step. Finally, the current performance of our prototype is well-suited for periodic audits, and further optimization would be necessary for its application in near-real-time trust-aware systems.

Future work will proceed along several key directions to address these limitations. We plan to evaluate the ACE on large-scale, anonymized enterprise datasets to validate its generalizability and real-world efficacy. To mitigate the policy formalization bottleneck, we will explore fully-automated techniques to the translation process. A key theoretical extension will be to enhance our model to handle correlated violations, where a single underlying fault may breach multiple rules, to ensure the resulting score accurately reflects the root cause without unfair penalization. By pursuing these extensions, we aim to build more transparent, accountable, and resilient trust ecosystems on the web.

%

%
%
\clearpage
\begin{credits}
\subsubsection{\ackname} This work was partially funded by the Horizon Europe project \href{https://datapact.eu/}{DATAPACT} (101189771) and the UKRI Horizon Europe guarantee funding scheme for the Horizon Europe projects \href{https://raise-science.eu/}{RAISE} (101058479) and \href{https://www.upcast-project.eu/}{UPCAST} (101093216).
\end{credits}

\bibliographystyle{splncs04}
\bibliography{mybibliography}

\clearpage
\appendix
\section{Illustrative Example of Semi-Automated Policy Formalization}
\label{sec:appendix_methodology_example}

This section provides a concrete, step-by-step walkthrough of our semi-automated, human-in-the-loop process for formalizing a compliance obligation. We use the HIPAA Authorization Control rule as a running example to demonstrate the workflow from a plain-text policy segment to a final, expert-verified logical rule. This process is designed to accelerate development while ensuring correctness and mitigating the risk of model-induced errors such as hallucination or logical misinterpretation.

\noindent
\textbf{Example: HIPAA Authorization Control (\texttt{hipaa\_auth})}

\noindent
\textbf{Step 1: Isolate Policy Text}

\noindent The process begins with a specific, actionable requirement from a compliance document.
\begin{quote}
\textit{"A principal acting in the role of a 'doctor' is only authorized to read a Patient Health Information (PHI) record if they are the designated physician for the patient who owns that record."}
\end{quote}

\noindent
\textbf{Step 2: Generate Candidate Rule with LLM}

\noindent This text is fed into a Large Language Model using a structured prompt that includes the schema definition and few-shot examples. The goal is to generate a candidate rule that captures the core logic. Below is a simplified representation of the prompt structure.

\noindent The LLM processes this prompt and generates the following candidate rule:

\begin{quote}
\textbf{LLM-Generated Candidate Rule:}\\
\small
\begin{verbatim}
violation('hipaa_auth', Doctor, PHI_Record) :-
    read_phi(Doctor, PHI_Record, _Purpose, _EventID),
    has_role(Doctor, 'doctor'),
    owns_phi_record(Patient, PHI_Record),
    \+ is_doctor_of(Doctor, Patient).
\end{verbatim}
\end{quote}

\noindent
\textbf{Step 3: Conduct Expert Verification and Finalization}

\noindent A human expert reviews the candidate rule for correctness, consistency, and safety.

\begin{itemize}
    \item \textbf{Logical Correctness:} The expert verifies that the logic correctly identifies a violation when a doctor is *not* the designated physician (`\ + is\_doctor\_of`). This accurately reflects the "if and only if" nature of the policy text.
    \item \textbf{Schema Consistency:} The expert confirms that all predicates (`read\_phi`, `has\_role`, etc.) are consistent with the established system schema, preventing ambiguity.
    \item \textbf{Error Mitigation:} The expert checks for potential LLM hallucinations or misinterpretations. In this case, the model correctly identified all necessary conditions without adding extraneous or incorrect logic.
\end{itemize}
The rule is confirmed as correct and is integrated into the final policy corpus. This same rigorous process is applied to all other policy segments.

\begin{tcolorbox}[colback=gray!5,colframe=black!60,title=\textbf{Structured Prompt Example},fonttitle=\bfseries]
\textbf{INSTRUCTION:} \\
Translate the following policy text into a declarative Prolog rule.
The rule should define a \texttt{violation} that occurs when a required 
constraint is not met after a trigger event.

\textbf{SCHEMA:}

\begin{verbatim}
- Predicates: read_phi(Principal, Resource, 
Purpose, EventID), 
  has_role(Principal, Role), 
  owns_phi_record(Patient, Resource), 
  is_doctor_of(Doctor, Patient), ...
- Rule Head: violation(RuleName, Subject, Object).
- Use '\+' for negation.
\end{verbatim}

\textbf{FEW-SHOT EXAMPLE:}
\begin{quote}\ttfamily
Text: "A patient's record shall not be processed for a purpose if they have not given consent." \\
Rule: \\
violation('gdpr\_art18\_restriction', P, R) :- \\
\quad read\_phi(P, R, Purpose, \_), \\
\quad owns\_phi\_record(Patient, R), \\
\quad \textbackslash+ has\_unrestricted\_status(Patient, Purpose).
\end{quote}

\textbf{POLICY TEXT TO TRANSLATE:}
\begin{quote}
"A principal acting in the role of a 'doctor' is only authorized to read a Patient Health Information (PHI) record if they are the designated physician for the patient who owns that record."
\end{quote}
\end{tcolorbox}

\section{Formal Policy Examples}
\label{sec:appendix_rules}

This appendix demonstrates the formalization of key compliance obligations from GDPR and HIPAA into the declarative Prolog rules used by our ACE auditor. Each rule defines the specific conditions that constitute a violation.

\noindent
\textbf{Rule 1: HIPAA Authorization Control (\texttt{hipaa\_auth})}
A violation occurs if a principal with a 'doctor' role reads a PHI record belonging to a patient to whom they are not formally assigned. This enforces the core principle of authorized access.

\begin{quote}
\small
\begin{verbatim}
violation('hipaa_auth', Doctor, PHI_Record) :-
    read_phi(Doctor, PHI_Record, _Purpose, _EventID),
    has_role(Doctor, 'doctor'),
    owns_phi_record(Patient, PHI_Record),
    \+ is_doctor_of(Doctor, Patient).
\end{verbatim}
\end{quote}

\noindent
\textbf{Rule 2: HIPAA Minimum Necessary (\texttt{hipaa\_min\_necessary})}
This rule enforces that a principal may only access the specific types of data (attributes) necessary for their role. A violation is triggered if a principal reads a record containing an attribute that their assigned role is not permitted to access.

\begin{quote}
\small
\begin{verbatim}
violation('hipaa_min_necessary', Principal, PHI_Record) :-
    read_phi(Principal, PHI_Record, _Purpose, _EventID),
    has_role(Principal, Role),
    read_attribute(Principal, PHI_Record, Attribute),
    \+ role_can_access_type(Role, Attribute).
\end{verbatim}
\end{quote}

\noindent
\textbf{Rule 3: GDPR Art. 18, Restriction of Processing (\texttt{gdpr\_art18\_ restriction})}
A violation occurs if a patient's record is processed for a specific purpose for which the patient has not given unrestricted consent. This enforces the data subject's right to control how their data is used.

\begin{quote}
\small
\begin{verbatim}
violation('gdpr_art18_restriction', Principal, PHI_Record) :-
    read_phi(Principal, PHI_Record, Purpose, _EventID),
    owns_phi_record(Patient, PHI_Record),
    \+ has_unrestricted_status(Patient, Purpose).
\end{verbatim}
\end{quote}

\noindent
\textbf{Rule 4: GDPR Art. 17, Right to Erasure (\texttt{gdpr\_art17\_erasure})}
This rule enforces the "right to be forgotten." A violation is detected if a patient's deactivation request remains unfulfilled for more than 30 days. The violation is attributed to the patient's assigned doctor, who is responsible for actioning the request.

\begin{quote}
\small
\begin{verbatim}
violation('gdpr_art17_erasure', Doctor, RequestID) :-
    request_deactivation(Patient, RequestID, RequestDate),
    is_doctor_of(Doctor, Patient),
    current_date(Today),
    days_since(RequestDate, Today, Days),
    Days > 30,
    \+ deactivation_fulfilled(RequestID).
\end{verbatim}
\end{quote}

\noindent
\textbf{Rule 5: GDPR Art. 15, Right of Access (\texttt{gdpr\_art15\_access})}
This rule upholds the patient's right to access their data in a timely manner. A violation occurs if a patient's formal request for their data is not fulfilled within the 30-day statutory limit. The patient's doctor is held responsible for this failure.

\begin{quote}
\small
\begin{verbatim}
violation('gdpr_art15_access', Doctor, RequestID) :-
    request_access(Patient, _PHI_Record, RequestID, RequestDate),
    is_doctor_of(Doctor, Patient),
    current_date(Today),
    days_since(RequestDate, Today, Days),
    Days > 30,
    \+ request_fulfilled(RequestID).
\end{verbatim}
\end{quote}

\section{System Parameters}
\label{sec:appendix_system_parameters}

To demonstrate the adaptability of the ACE framework to diverse organizational risk profiles, Tab.~\ref{tab:system_parameters} provides a list of the system's tunable parameters. These configuration options allow administrators to calibrate the scoring model's sensitivity, dimension prioritization, and temporal dynamics.

\begin{table}[htbp]
\centering
\caption{Comprehensive List of System Parameters}
\label{tab:system_parameters}
\small
\begin{tabular}{p{1.8cm} p{3.5cm} @{\hspace{0.3cm}} p{6.5cm}}
\toprule
\textbf{Parameter} & \textbf{Description} & \textbf{Impact on Reputation Computation} \\ \midrule

Time \par  Window \par ($W$) & 
The discrete time period (e.g., weekly, monthly) over which violations are grouped and assessed. & 
Determines the update frequency of the score. A shorter window provides near real-time volatility; a longer window smoothes out behavioral spikes. \\ \addlinespace

Rule \par Criticality \par ($C$) & 
A pre-assigned weight $C \in [0,1]$ reflecting the intrinsic importance of a specific compliance rule. & 
Prioritizes high-risk obligations. A high $C$ (e.g., 0.95 for GDPR Erasure) ensures that a single violation significantly degrades the score, whereas low $C$ violations have minimal impact. \\ \addlinespace

Scaling \par Factor \par ($\alpha$) & 
A scaling parameter unique to each dimension ($\alpha_V, \alpha_T, \alpha_B$) used in the normalization function. & 
Controls the system's sensitivity to violation magnitude. A small $\alpha$ creates a "strict" system where even minor violations result in high severity scores. A large $\alpha$ creates a "lenient" system requiring massive violations to trigger severe penalties. \\ \addlinespace

Dimension \par Weights \par ($w$) & 
The relative weights ($w_V, w_T, w_B$) assigned to Volume, Duration, and Breadth in the geometric mean calculation. & 
Tailors the score to organizational priorities. For example, increasing $w_V$ makes the score highly sensitive to massive data leaks, while increasing $w_T$ penalizes persistent, long-term non-compliance more heavily. \\ \addlinespace

Decay \par Constant \par ($\lambda$) & 
A time-decay factor determining how quickly historical penalties are reduced over time. & 
Controls the "path to redemption." A high $\lambda$ allows the score to recover quickly after violations cease (high forgiveness). A low $\lambda$ causes penalties to linger, requiring a long period of compliant behavior to restore the score. \\ \bottomrule

\end{tabular}
\end{table}
\end{document}